\documentclass[aps,twocolumn,prl,superscriptaddress,showpacs]{revtex4}

\usepackage{latexsym,amsmath,verbatim,amsfonts}
\usepackage{graphicx}

\newcommand{\av}[1]{\langle {#1} \rangle}

\begin{document}

\title{Dynamical and bursty interactions in social networks}

\author{Juliette Stehl\'e} 
\affiliation{Centre de Physique Th\'eorique (CNRS UMR 6207), Luminy,
13288 Marseille Cedex 9, France}

\author{Alain Barrat}
\affiliation{Centre de Physique Th\'eorique (CNRS UMR 6207), Luminy,
13288 Marseille Cedex 9, France}
\affiliation{Complex Networks and Systems Group,
Institute for Scientific Interchange (ISI), Torino, Italy}

\author{Ginestra Bianconi}
\affiliation{Department of Physics, Northeastern University, Boston, 02115, MA,USA}
\date{\today}

\begin{abstract}
  We present a modeling framework for dynamical and bursty contact
  networks made of agents in social interaction. We consider agents'
  behavior at short time scales, in which the contact network is formed by disconnected
  cliques of different sizes.  At each time a random agent can make a
  transition from being isolated to being part of a group, or
  viceversa.  Different distributions of contact times and
  inter-contact times between individuals are obtained by considering
  transition probabilities with memory effects, i.e. the transition
  probabilities for each agent depend both on its state (isolated or
  interacting) and on the time elapsed since the last change of state.
  The model lends itself to analytical and numerical investigations.
  The modeling framework can be easily extended, and paves the way for
  systematic investigations of dynamical processes occurring on
  rapidly evolving dynamical networks, such as the propagation of an
  information, or spreading of diseases.
\end{abstract}



\maketitle


Recently, technological advances have made possible the measure of
social interactions in groups of individuals, 
at several temporal and
spatial scales and resolutions showing that human activity obeys scaling law and statistical features
which reveal long time correlations and memory effects.
Evidence comes from  data on email exchanges \cite{Eckmann:2004,Barabasi:2005,Havlin2:2009},
mobile phone communications \cite{Onnela:2007,Hidalgo:2008}, 
spatial proximity \cite{Hui:2005,Eagle:2006,Kostakos,Pentland:2008}, web browsing \cite{Vazquez:2006}
and even face to face interaction
\cite{Sociopatterns,Sociopatterns2,Sociopatterns3}.
In this respect the traditional framework of models used for risk assessment and communication,
which describe human actvity as a series of Poisson distributed processes, need to be changed in favor of
new models which take into account 
the occurrence of burstiness in many aspects of human activity.

Social interactions give rise to social
\cite{Granovetter:1973,Wasserman:1994} and collaborative
\cite{Newman:2001} networks characterized by a complex evolution.  In
these networks, links are constantly created or terminated, and the
social network of an individual evolves at different levels of
organization.  After the pioneering papers on complex networks showing
that many social networks are small world and display heterogeneous
degree distributions \cite{reviews}, and that these network topologies
strongly influence the dynamics taking place on the
networks \cite{Barrat:2008}, a number of papers have been devoted to 
modeling the dynamics of social interactions. Issues investigated in this context are in
particular community formation \cite{Kumpula:2007,Johnson:2009} and
the evolution of adaptive dynamics of opinions and social ties
\cite{Bornholdt:2002,Marsili:2004,Holme:2006,MaxiSanMiguel:2008}.

The evidence coming from the analysis of social contact data calls for
new frameworks that integrate these models with the bursty character
of social interactions. The duration of contacts between individuals
or groups of individuals display indeed broad distributions, as well
as the time intervals between successive contacts
\cite{Hui:2005,Scherrer:2008,Sociopatterns,Sociopatterns2}.  Such
heterogeneous behaviors have strong consequences on dynamical
processes \cite{Vazquez:2007,Onnela:2007}, and should therefore be
correctly taken into account.  It is therefore necessary to introduce
this fundamental aspect on human activity in models of social
interactions, possibly reconstructing then social networks by
aggregating the network of contacts over a certain period
\cite{Holme:2005,Vazquez:2007,Havlin:2009}. The modeling literature in
this area being still in its infancy
\cite{Gross:2008,Scherrer:2008,Hill:2009,Gautreau:2009,Latora:2009},
it is important to develop simple, generic, easily implementable
models of dynamical networks which reproduce the empirical facts
observed in contact duration and inter-contact intervals.

In this Letter, we take a step in this direction, focusing on short
timescales such as the ones involved when people interact in social
gatherings (e.g., scientific conferences). We define a simple
agent-based model for rapidly evolving sparse dynamical networks,
aimed at describing the dynamics of human social interactions in the
context of small discussion groups.
In particular we are interested in
investigating basic mechanisms which could be responsible for various
contact duration distributions. The model is kept simple, so that it
can be easily simulated. It is accessible to analytical investigations
in a certain number of cases. It can also be easily extended or
modified. For instance, the population of agents is considered
homogeneous (i.e. every agent is assumed to have the same dynamical
parameters) and an extension to heterogeneous populations can easily
be envisioned.

The dynamical network under study is formed by disconnected
groups of agents which
evolve by successive mergings and splittings. In particular at each
time step an agent can either leave or remain in its group, or
introduce an isolated agent to its group.  The general formulation of
the model allows to describe a variety of behaviors of the dynamical
networks. In particular, the duration of contacts between individuals
can display either narrow or broad distributions. A narrow
distribution is for instance obtained by simply assuming that each
agent leaves a group or invites a new agent in its group with a
time-independent probability. On the contrary, broad distributions of
contact durations, similar to those observed in empirical studies
\cite{Hui:2005,Eagle:2006,Kostakos,Scherrer:2008,Sociopatterns}, are
obtained through a reinforcement dynamics of the interaction, that can
be summarized as {\it ''the longer an agent interacts with a group,
the less it is likely to leave the group;
the more the agent is isolated the less likely it is to interact with a group''}.  This dynamics,
reminiscent of the preferential attachment in the context of complex
networks \cite{Barabasi:1999a}, could be argued to stem from
Hebbian-like mechanisms at the underlying cognitive level. In general,
for both narrow and broad distributions of interaction times, larger
groups are found to be less stable than smaller ones. This is also
observed in the data \cite{Sociopatterns} and can be simply explained:
the lifetime of a group depends on the decisions of all its
members. In a first approximation these decisions correspond to
independent events, therefore groups with more agents become less
stable. Interestingly, our model also exhibits a dynamical transition
towards the formation of large size group. This transition, supported
by some measurements in animal behavior \cite{Morgan:1976,Levin:1995},
is not observed in human behavior, and corresponds thus to
parameter values where the model loses its applicability 
to the description of human social interactions.
Note, nevertheless, that the
formation of large social organizations and cities, demonstrate that
in humans, large group formation occurs at a different level of
organization.


The model we propose considers a fixed population of $N$ agents,
interacting in a limited space, as for example in a
conference venue
\cite{Pentland:2008,Kostakos,Sociopatterns,Sociopatterns2,Sociopatterns3}.
Therefore, in a first approximation we neglect the spatial
dispersion of the agents and assume a well mixing dynamics. 
Each agent can either be isolated or belong to a group
with other agents, and the groups define an instantaneous contact
network. During the dynamics, agents can join other agents or on the
contrary leave the group they belong to. More precisely, each agent
$i$ is characterized by two variables: the number $p_i$ of other
agents with which it is in contact (i.e. its degree in the
network) and the time $t_i$ at which $p_i$ last evolved.  At each time
step $t$, an agent $i$ is chosen at random. If $i$ is isolated
($p_i=0$), $i$ changes its state with probability $b_0 f(t,t_i)$.  
In this case, another isolated agent $j$ is chosen with a
certain probability $\Pi(t,t_j)$, and $i$ and $j$ form a pair ($p_i
\to 1$, $p_j \to 1$ and $t_i\rightarrow t$, $t_j\rightarrow t$).
If on the other hand $i$ is part of a group ${\cal G}$ of size greater
than one (i.e. $i$ has degree $p>0$), a change of state occurs with
probability $b_1 f(t,t_i)$. When this occurs, agent $i$ can either
leave the group (probability $\lambda$) or introduce an isolated agent
in the group (probability $1-\lambda$).  If $i$ leaves the group
${\cal G}$ and becomes isolated, $p_i \to 0$, and $p_j \to p -1\,
\forall j\in {\cal G}\setminus\{i\}$, and as a consequence of this
event the time of the modified nodes is reset to $t$,
i.e. $t_{\ell}\rightarrow t\, \forall {\ell}\in{\cal G}$. If $i$
introduces to the group an isolated agent $j$, chosen again with
probability $\Pi(t,t_j)$, then $p_{\ell} \to p+1 \, \forall
\ell\in{\cal G}\cup\{j\}$ and each agent $\ell$ in ${\cal G}\cup
\{j\}$, changing state at
time $t$, sets $t_\ell\rightarrow t$.
The parameters $b_0$ and $b_1$ determine the tendency of the agents,
respectively isolated or in a group, to change their state, while
$\lambda$ controls the tendency either to leave groups or on the
contrary to make them grow. The model's dynamical behavior depends
also on the functions $f$ and $\Pi$. 

In order to make contact with empirical data, the main quantities of
interest concern the time spent by agents in each state, the duration
of contacts between two agents, and the time intervals between
successive contacts of an agent. We can gain insight into these
properties by  writing rate equations for the
evolution of the number $N_p(t,t_0)$ of agents which are at time $t$
in state $p$ since $t_0$:
\begin{eqnarray}
\nonumber
\partial_t N_0(t,t_0) &=& -\frac{N_0(t,t_0)}{N} b_0 
\left[ f(t,t_0) \right. \\ \nonumber
&+& \left. \Pi(t,t_0) (r(t) + (1-\lambda) \alpha(t) )
\right]
+ \sum_{p \ge 1} \pi_{p,0}(t) \delta_{t,t_0} \\ \nonumber
\partial_t N_1(t,t_0) &=& -2 \frac{N_1(t,t_0)}{N} b_1 f(t,t_0) + 
( \pi_{0,1}(t) + \pi_{2,1}(t) ) \delta_{t,t_0} \\ \nonumber
\partial_t N_p(t,t_0) &=& - (p+1) \frac{N_p(t,t_0)}{N} b_1 f(t,t_0) \\ \nonumber
&+& \left( \pi_{p-1,p}(t) + \pi_{p+1,p}(t) + \pi_{0,p}(t) \right) \delta_{t,t_0} 
,\ p > 1
\end{eqnarray}
where $\pi_{p,q}(t)$ is the average number of agents going from state
$p$ to state $q$ at time $t$, and
\begin{equation}
r(t) = \frac{\sum_{t'}N_0(t,t') f(t,t')}{\sum_{t'}N_0(t,t') \Pi(t,t')}
\end{equation}
\begin{equation}
\alpha(t)= \frac{\sum_{p\geq 1,t'}N_p(t,t')b_1 f(t,t')}{\sum_{t'}N_0(t,t') b_0
\Pi(t,t')}
\end{equation}
where in the sums $t' < t$.
These equations can be simplified and solved in certain
cases, and the distribution $P_p(\tau)$ of (normalized) times
$\tau=(t-t_0)/N$ in which an agent remains in a given state $p$ can
then be deduced. Let us for definiteness assume that $f$ and $\Pi$ are
stationary functions so they depend only on $t-t_0$; it is then natural to look for a stationary
state, reached at large enough times, such that $\alpha$, $r$,
$(\pi_{p,q})$ are constants, and $N_p(t,t_0)=N_p(t-t_0)$.  If for
instance $f$ is a constant, it is easy to see that the $\left\{N_p\right\}_{p \in \mathbb{N}}$ decay
exponentially with time, so that the $\left\{P_p(\tau)\right\}_{p > 0}$ are as well
exponentially decaying functions.

We consider the more interesting case of $f$ and $\Pi$ {\em decaying}
with $t-t_0$: the more an agent is in a state, the less probable it
becomes to change state, as previously described in the
self-reinforcement mechanism. For sake of simplicity, we focus on the
case $f=\Pi$, so that $r=1$, which allows to simplify the computations
\footnote{In general, one sees that $P_p(\tau)$ for $p>0$ depends only
  on $f$ and $b_1$, while $P_0$ depends on both $f$ and $\Pi$.}.
Computations can be carried out completely for instance in the case
$f(t-t_0)= \Pi(t-t_0)= (1+\tau)^{-1}$, where $\tau=(t-t_0)/N$. The
choice of this scaling is consistent with the scaling of email
communications and other human activity
\cite{Barabasi:2005,Vazquez:2006}.  In particular, $N_p(t)$ for $p \ge
1$ is readily seen to decay as a power-law with exponent $(p+1)b_1$.
More involved computations are needed to obtain the decay exponent of
$N_0$. Writing $\left\{\pi_{p,q}\right\}$ as functions of
$\left\{N_p\right\}_{p \in \mathbb{N}}$, $f$ and $\Pi$, we can obtain
recurrence relations for $\left(\pi_{p,0}\right)_{p > 0}$ and deduce
$\alpha=(2\lambda -1)^{-1}$, so that
\begin{eqnarray}
P_0(\tau) &\propto& (1+\tau)^{-1+b_0\frac{3\lambda-1}{2\lambda-1}} \\
P_p(\tau) &\propto& (1+\tau)^{-1+(p+1)b_1} \ , \ p \ge 1 \ .
\end{eqnarray}
The previous analytical results are obtained under the conditions
$b_1 > 1/2$, $\lambda > 1/2$, $b_0 > (2\lambda-1)/(3\lambda-1)$, which
determine the phase diagram of the model: outside these boundaries, the hypothesis
of stationarity is violated.

%
%
\begin{figure}[t]
\begin{center}
\includegraphics[width=0.45\textwidth]{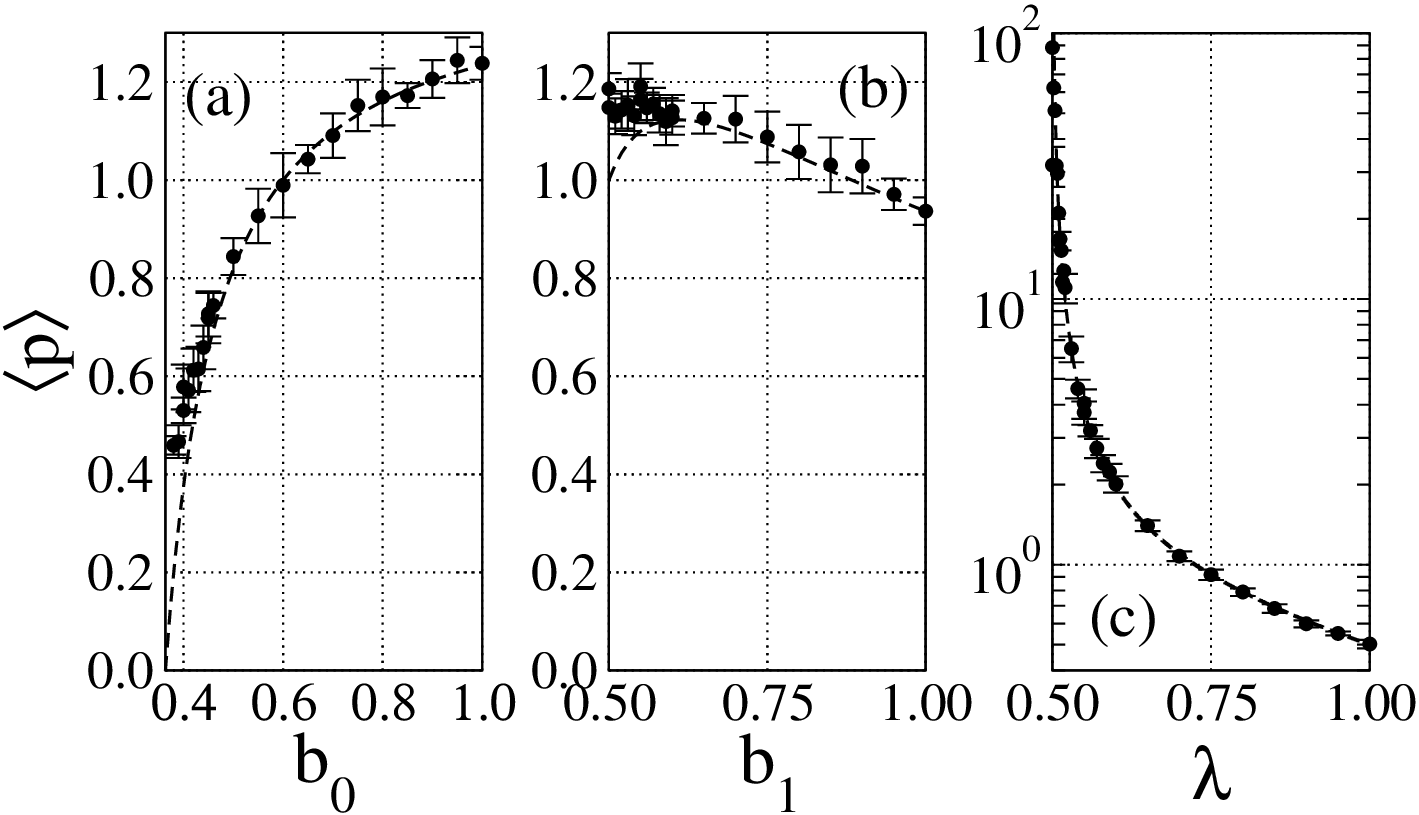}
\caption{Average state $\av{p}$ in the stationary state for different sets of parameters and a population 
of $N=2000$ agents. 
(a) $b_1=\lambda=0.7$; (b) $b_0=\lambda=0.7$; (c) $b_0=b_1=0.7$. In (c) we observe the divergence of
the average group size as $\lambda \rightarrow 0.5^{+}$. The lines show Eq. (\ref{averagestate}).}
\label{fig:averagestate}
\end{center}
\end{figure}
It is also possible to compute the average state of the agents
in the stationary state, as
\begin{equation}
  \langle p \rangle = \frac{\pi_{1,0}}{2 \lambda} 
\sum_{i \geq 1}\frac{i(i+1)}{(i+1)b_1 -1} 
\left(\frac{1-\lambda}{\lambda} \right) ^{i-1}
\label{averagestate}
\end{equation}
where
\begin{equation}
 \label{pi} \nonumber
\pi_{1,0} = \left[ 
\frac{1}{2 \left( b_0 - \frac{2 \lambda -1}{3 \lambda -1}\right)}+ 
\frac{1}{2 \lambda} \sum_{i \geq 2} \frac{i}{i b_1 -1} 
\left(\frac{1-\lambda}{\lambda}\right)^{i-2}\right]^{-1} .
\end{equation}
For $\lambda\rightarrow 0.5^{+}$
the average group size $\av{p}$ diverges indicating that, in this
limit, the non-stationary state is dominated by the formation of a
large group of size ${\cal O}(N)$.

\begin{figure}[t]
\begin{center}
\includegraphics[width=0.45\textwidth]{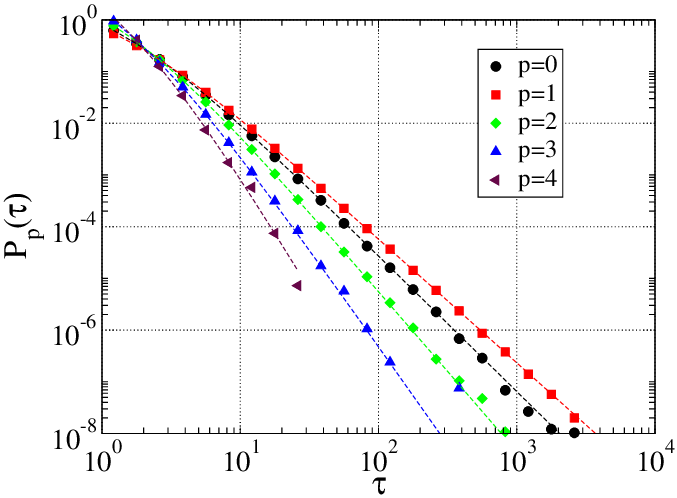}
\end{center}
\caption{Distribution $P_p(\tau)$ of times during which an agent
  does not change connectivity $p$. $N=10000$, $b_0=b_1=0.7$, $\lambda=0.8$, and the
simulation is run for  $T=10^5 N$ time steps. The lines are the analytical predictions. }
\label{fig:ptau_ness}
\end{figure}

We have performed numerical simulations of dynamical networks generated by the
present model, with different $f$, $\Pi$, values of the parameters
$b_0$, $b_1$, $\lambda$, and sizes $N$. We will here show the simulations
corresponding to $f=\Pi=(1+\tau)^{-1}$, in order to compare with
the analytical predictions presented above. 
We first show in Fig.~\ref{fig:averagestate} the average agent state 
as a function of the different parameters, recovering the behavior
predicted in Eq.(\ref{averagestate}). The average state 
increases with $b_0$, decreases with $\lambda$, and presents a 
non-monotonous behavior with $b_1$.
Figure \ref{fig:ptau_ness} displays the distributions
$\left\{P_p\right\}_{0\leq p \leq 4}$ of time
spent in the various states. These distributions are power-laws, in
perfect agreement with the analytics. We also note that, for
$f=\Pi=(1+\tau)^{-\nu}$ with $\nu \ne 1$, $\left\{P_p(\tau)\right\}_{p
  \in \mathbb{N}}$ can be
shown analytically to become either stretched exponentials ($\nu<1$)
or power-laws ($\nu>1$), and we have also checked this behavior numerically.
The broadness of the distributions is therefore not limited to the particular case
described above, but is quite robust with respect to changes in the
microscopic rules.

\begin{figure}[t]
\begin{center}
\includegraphics[width=0.45\textwidth]{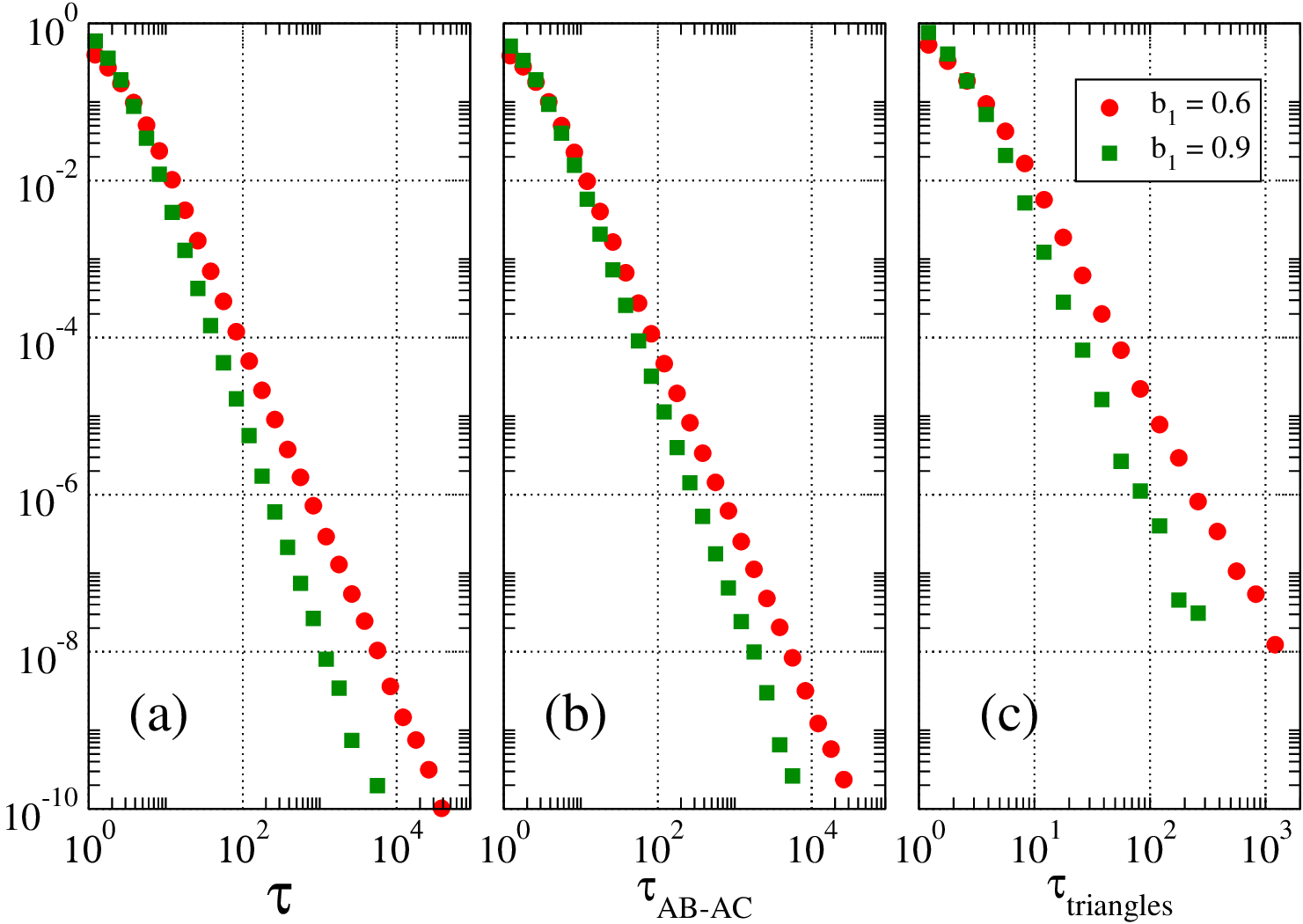}
\end{center}
\caption{Distributions of (a) duration of a contact between two agents; (b) 
time intervals between the beginnings of successive contacts of an agent A with two different agents B and C;
(c) duration of a triangle.
$b_0=0.7$, $\lambda=0.8$, $b_1=0.6$ and $0.9$. $N=1000$, $T=10^5N$.
}
\label{fig:contacts}
\end{figure}

In Fig. \ref{fig:contacts}, we also show the distribution of contact
durations between two agents (which is different from $P_1$: two
agents remain in contact when they are joined by a third, but leave
the state $p=1$), of triangle durations, and
of the time intervals between the starting times of
two successive contacts \cite{Sociopatterns}. This last quantity is
highly relevant in the context of causal processes, as it gives the
time scale on which an agent can propagate an information or a disease
after receiving it. All these distributions are broad,
similarly to empirical observations \cite{Scherrer:2008,Sociopatterns}.

Let us finally mention that, when considering parameter values outside the validity of
the stationary state analytical computations, different scenarios are
observed, depending on $\lambda$: if $\lambda > 0.5$, the average
state slowly decreases (towards $0$ if $b_0<0.5$, and $1$ if $b_1<0.5$)
while, for $\lambda < 0.5$, a large cluster appears, with size is proportional to
$N$, and lasting on a diverging timescale.
Interestingly, even in
this non-stationary case, the shape of the distributions
$P_p(\tau)$ may remain stationary (not shown).
This is particularly relevant as most
empirical data are necessarily obtained in non-stationary
environments.

In this Letter, we have proposed a modeling framework for dynamical
networks in the context of interacting social agents.  Both broad and
narrow distributions can be obtained, corresponding to different
social situations. The present framework can be developed in several
research directions. First, many variations of the microscopic rules
may be thought of and implemented, in order to model more precisely
mechanisms of social contacts in various contexts or even of
animal behavior. For instance, merging and splitting of groups could be
introduced, as well as heterogeneity between agents to take
into account different propensities
to interact or to create groups. Moreover, it will be interesting to
investigate how the properties of the interaction durations shape the
resulting aggregated networks on various timescales. 
Finally, model dynamical contact networks can be used as a support for the
simulation of dynamical processes taking place on dynamical networks, such
as information spreading in a conference: the spreading process, although
taking place on an extremely sparse network which is at any time 
formed of disconnected groups, may overall concern the whole population
of agents, thanks to the dynamics of the agents who move from one
group to another.
The fact that the various network characteristics (such as the
broadness of the distribution of contact durations and inter-contact times)
can be controlled by changing the model's parameters will then make
it possible to understand better the effect of these characteristics
on the dynamical processes under scrutiny.

\textit{Acknowledgments} 
This work has been partly supported by the FET Open project DYNANETS
(number 233847) funded by the European Commission.

\end{document}